# Enhanced capillary pumping through evaporation assisted leaf-mimicking micropumps

Prasoon Kumar[a,b,c], Prasanna S Gandhi[b], Mainak Majumder[c]

a IITB-Monash Research Academy, Powai, Mumbai, Maharashtra-400076, India

b Suman Mashruwala Advanced Microengineering Laboratory, Department of Mechanical Engineering, Indian Institute of Technology Bombay, Powai, Mumbai, Maharashtra 400076, India

c Nanoscale Science and Engineering Laboratory (NSEL), Department of Mechanical and Aerospace Engineering, Monash University, Clayton, Melbourne, Australia

## Abstract

Pumping fluids without an aid of an external power source are desirable in a number of applications ranging from a cooling of microelectronic circuits to Micro Total Analysis Systems (µ-TAS). Although, several microfluidic pumps exist, yet passive micropumps demonstrate better energy efficiency while providing a better control over a pumping rate and its operation. The fluid pumping rate and their easy maneuverability are critical in some applications; therefore, in the current work, we have developed a leaf-mimicking micropump that demonstrated ~6 fold increase in a volumetric pumping rate as compared to the micropumps having a single capillary fluid delivery system. We have discussed a simple, scalable, yet inexpensive method to design and fabricate these leaf mimicking micopump. The microstructure of the micropumps were characterised through scanning electron microscopy and its pumping performance (volumetric pumping rate and pressure head sustainence) were assessed experimentally. The working principle of the proposed micropump is attributed to its structural elements; where branched-shaped microchannels deliver the fluid acting like veins of leaves while the connected microporous support resembles mesophyll cells matrix that instantaneously transfers the delivered fluid by a capillary action to multiple pores mimicking the stomata for evaporation. Such design of micropumps will enable an efficient delivery of the desired volume of a fluid to any 2D/3D micro/nanofluidic devices used in an engineering and biological applications.

# 1. Introduction

The precise control over the volume of fluid at a given flow rate is advantageous in a number of applications like advanced drug delivery, cooling circuits in electronic devices, chromatographic column in chemistry and many more (1-3). The current solution to the above problems is addressed by micropumps. Therefore, several engineering principles have been employed to develop micropumps like electrostatic, magnetic, piezoelectric, thermo-pneumatic, electrochemical, electrohydrodynamic, electrokinetic, electro-osmotic and magneto-hydrodynamic (4). However, they are either power intensive or deliver oscillating/pulsating flows that are undesirable in certain conditions. Moreover, the volume of fluid delivered and its rate is significantly dependent on the power source. Therefore, capillary based fluid pumping available in trees can be emulated to structurally mimic the design and material properties for the development of passive micropumps (Figure 1a) (5, 6)

Several efforts have been made to design and develop pumps that can drive volume of fluid comparable to that in a tree in a controlled fashion. However, they had one or other shortcomings - either the pumping of water to a desired height was not possible or the flow rates were too low compared to that observed in natural trees. One of the best passive pumping heights achieved by an engineered mechanical system was up to 17m (7-10). Therefore, micropumps inspired from the transpiration assisted fluid pumping by leaves of trees have been pursued by recent researchers to harness pumping capabilities comparable to that of the trees.

The pioneering work on artificial trees, based on hydrogel systems, proposed by Wheedle *et. al.* reported a fluid pumping rate of 0.014 mg/sec. The proposed artificial tree comprised of two hydrogels; one representing the root and other the leaves, connected by microchannels that acted as the trunk of a tree. Their study was primarily directed towards the formation and effect of cavitation on the negative pressure that prevented further pumping of fluid in trees. Thus, hardly any attention was given to enhance the fluid pumping rate through modification in the design or the parameters affecting the fluid pumping behaviour in such artificial systems (11). Jingmin li *et. al.* fabricated a functional and structural mimic of leaves of plants to develop a micropump using agarose gel (resembling mesophyll cells of leaves) and silicon membrane having slit like micropores (mimic of leaf's stomata) along with other supportive components. These micropumps could emulate the embolism effect as observed in the xylem of natural plants under an extreme condition in addition to achieving the fluid pumping rate of 1.12ul/min. They also estimated the maximum water potential developed in their proposed micropump could be 72.5KPa, a pressure substantial to lift a water column to a height of 7m but they haven't experimentally verified it (12, 13). However, the silicon micropores were insensitive to temperature due to the fixed aperture size, leading to difficulty in controlling the fluid pumping rate. Therefore, Hung *et. al.* developed thermo-responsive polymer based porous membrane mimicking leaves' stomata and integrated with agarose gel based system to develop leaf-inspired micropump. It can emulate the closure of stomata with temperature variation, thereby controlling the pumping rate for different fluidic applications. They reported a flow rate of 30ug/min which is much higher than any previously reported data on fluid pumping rate (14) .

The above efforts were primarily driven towards creating a better mimic of leaves or trees by tailoring the design or material to emulate the pumping functions as observed in trees. These authors used hydrogels as materials to pump the fluids from a reservoir through a single capillary tube, to the microporous substrate for evaporation. The hydrogels were used due to their highly hydrophilic nature, high swelling capacity and good interconnected microporous network. However, application of hydrogels may not be an optimal strategy to deliver fluid to microporous substrate to support evaporative pumping due to several reasons. First, hydrogels are structurally weak materials, primarily used in soft tissue engineering applications. (15). Hence, they are least suitable for applications demanding larger sized hydrogels during scaling up of a pumping device. Even if they need to be used, these hydrogel requires additional support structures that will unnecessarily complicates the design of the micropumps. Second, these hydrogels swells in the presence of water, hence care has to be taken to prevent delamination of microporous structure participating in evaporation from the hydrogels due to different swelling ratio (12, 14) and thirdly, hydrogels have random network of inter-connected pores that may increase the hydraulic resistance during capillary flow in applications demanding suction head. Eventually, hydrogels based micropumps are expensive to fabricate, involves complicated design and have limited applicability.

The above limitations of hydrogels led Jingmin li *et al* to explore microporous membrane on a planner substrate connected to a microchannel as one of the simplistic models of micropump for mimicking transpiration-assisted capillary pumping. The microporous membranes overcome majority of limitations of hydrogels listed above while developing passive micropumps. The reported micropump achieved the controlled flow rate within 0.13–3.74ul/min through changes in intrinsic and ambient parameters. The parameters they considered were temperature, humidity, size and number of micropores in silicon membranes participating in transpiration. Later, the few intrinsic parameters were studied in detail by Robert Crawford et al. They suggested that the length and diameter of a channel irrigating the microporous membrane and area of the membrane participating in fluid pumping as an essential design criteria for leaf-inspired micropumps. They could neither observe any appreciable change in fluid pumping rate nor in the attained suction head due to changes in the diameter of the microchannel irrigating the microporous membrane. However, the area of porous membrane and pore size was found to have profound impact on the pumping rate of leaf-inspired micropumps. They further commented that single capillary delivering fluid to microporous substrate may not be sufficient to irrigate its entire surface. Hence, this may have led to under-utilization of the available surface during evaporation pumping at a given temperature, humidity and permeability of the microporous membrane (16). The above findings demonstrated the dependence of performance of leaf-inspired micropumps on the microchannel irrigating the microporous membrane under a given set of ambient parameters. Further, it has been known that the branching channel network supports efficient fluid flow with least flow resistance and they are the usual venation system for irrigating the non-swelling spongy cells of leaves for supporting evaporation dependent pumping in trees(17).

Thus, inspired from the design of leaves, we hypothesize that developing a leaf-mimicking micropump having a branching channel network with multiple fluid delivery points for irrigating microporous substrate will lead to an enhanced fluid pumping through evaporation. In the present work, we proposed a simple design of leaf-mimicking micropump (LMM) that is capable of achieving high fluid pumping rate and suction head maintenance. The LMM was fabricated through a simple, inexpensive and scalable method. The proposed LMM comprises of radially arranged, branch shaped, multi-scale microchannels in Polydimethysiloxane (PDMS) (representing venation system of leaves) and a non-swelling, microporous substrate (resembling stomata and spongy mesophyll cells of leaves). The microstructure of LMM were characterised through SEM and was evaluated for pumping performance. Thus, our work is expected to enable the fabrication of efficient LMMs with tailored geometric properties suited for different fluid pumping applications.

## 2. Materials and Methods

### 2.1. Materials

Microporous cellulose filter paper (Whattman, 90mm Diameter) and Polydimethylsiloxane (PDMS) (Sylgard® 184 silicone *elastomer* kit - *Dow* Corning) was used for the fabrication of leaf-mimicking micropump device. The photo-resist ceramic suspension was prepared from a monomer HDDA (1, 6 Hexanediol diacrylate) (Sigma Aldrich) as per the recipe described by Tanveer et al (18). Deionised (DI) water was used as a working fluid; Disodium Fluorescein (Sigma Aldrich Pvt. Ltd., India) was used as a dye and polytetrafluoroethylene (PTFE) tubes were used for tubing and connections.

### 2.2. Methods

#### 2.2.1. Fabrication process of leaf-mimicking micropumps (LMM)

The fabrication process of leaf-mimicking micropump (LMM) is shown in the Figure 2. The branching microstructures were fabricated with thoroughly mixed ceramic suspensions in HDDA (18). A drop of ceramic suspension was placed on a glass slide and inserted in a controlled Hele-Shaw apparatus indigenously developed in the lab (18, 19). The separation of glass slides as shown in the Figure 2(a) result in the formation of a branching microstructures. These obtained structures were heated over a hot plate at 120°C for a period of 24 hours to cure and form a stable mold ready for casting. Further, PDMS solution was prepared by thorough mixing of base and curing agent in 10:1 ratio and degassed before being poured over the prepared ceramic mold. Thereafter, PDMS was cured over a hot plate at 70°C for 6-8 hours. Then, it was cooled at a room temperature and carefully peeled off from the mold to form open branched shaped microchannels network in PDMS block having a thickness of 2mm. In parallel, PDMS was spin coated at 800rpm for 2-3 minutes on a paraffin wax paper. It was left at room temperature for 24 hour to get it partially cured. Thereafter, sticky PDMS film on a paraffin wax paper was placed onto microporous filter paper and pressed to allow adherence of PDMS film with microporous filter paper. Thereafter, the paraffin wax paper was carefully removed and the open branched shaped microchannels in PDMS block were sealed with the PDMS film adhered to the microporous filter paper such that the terminal microchannels were connected to the microporous paper. The LMM device as shown in Figure 2(f) was heated at 60 °C for 1-2 hours to strengthen the bonding between PDMS block with branched shaped microchannels and microporous filter paper by an intermediate PDMS film. The control device was fabricated by bonding a PDMS block having no microchannel net with a microporous filter paper.

**2.2.2. Structural Characterization of LMMs**

The mold of branch-shaped microstructures was produced by controlled Hele-Shaw apparatus, and branched shaped microchannels in PDMS block were taken for structural characterization after imaging by a camera (Sony). The structural parameters of the samples were analysed and reported after image processing in MATLAB® 2009a. Further, cross-section of micropump's structures and microporous support were taken for morphological characterization by scanning electron microscopy (SEM). The samples were mounted onto copper stubs and prepared with a gold coating using a sputter coater for scanning electron microscopy (SEM) (FEI QUANTA 200, FEI, USA,), operated at 20 kV and 10 kV, respectively. The images were taken at different magnifications for a complete morphological analysis. Further, white light interferometry module of Polytec MSA 500 (Micro System analyzer) was used for surface profilometry to estimate the height of the features of branched shaped microstructures. The microporous substrate was evaluated for volume of water absorbed when saturated to determine its absorption capacity. Further, the porosity of microporous substrate was evaluated by gravimetric method (20) and given by formula

$$\% \; porosity = \left(1 - \frac{\rho_s}{\rho_m}\right) \times 100 \qquad (1)$$

where $\rho_s$ is the density of microporous substrate and $\rho_m$ is the density of material of microporous substrate.

### 2.2.3. Study of fluid pumping rate and pressure head by the LMMs

The LMM fabricated above were connected at an entrance of parent channel to one end of a PTFE tube (length 79cm and internal diameter 2mm) while the other end of the tube was connected to a reservoir as shown in figure 3. Initially, the reservoir and LMM was placed at a same level from the ground and thereafter, slowly reservoir level was raised above the LMM by 2mm to enable water to flow through the pump. The flow of water from the reservoir was continued till the entire microporous paper part of the LMM was well irrigated to a saturation level. Thereafter, the reservoir was steadily lowered till the level of LMM and an air bubble was introduced at an entrance of the tube where it was dipped in a reservoir. This raising the level of reservoir in comparison to LMM triggers the flow of water from reservoir to the pump via the connecting tube. The LMM was left in an open air in a clean room environment at 25$^{\circ}$C and relative humidity of 50%. Then, the fluid pumping velocity was measured by tracking the movement of air bubble in the tube when reservoir and LMM was placed at a same height. Thereafter, the platform on which LMM was placed was continuously, slowly raised in steps. At each step, the meniscus of introduced air bubble was followed to track the flow of water in the tube from reservoir to the LMM for period of 15 minutes. The process of raising the LMM was ceased when there was no appreciable change in the position of air bubble was observed in the tube. Control experiments were carried out with the microporous paper connected to the reservoir via a tube without any intermediate branched-shaped microchannel networks between porous support and tube. The volumetric flow rate under different pressure head was calculated from the above recorded data.

## 3. Results and Discussion

The current paper elucidated a process of design and fabrication of biological leaf-mimicking micropumps. These micropumps emulated the structural design of natural leaves at a different length scale to achieve fluid pumping capabilities. It comprised of multi-scale, branched-shaped microchannel networks in PDMS, mimicking venation system of leaves, to irrigate microporous paper (resembling mesophyll spongy cells having stomata) with a fluid (Figure 1b). However, the transpiration assisted micropumps described in literatures (11-14, 16) were primarily microporous membranes or hydrogels irrigated by a single microchannel. The hydrogels outperformed microporous membranes for fluid pumping capacity due to their high water swelling capacity. However, during evaporative pumping, the pumping capacity is dependent on the surface area available for evaporation. Due to structurally weak material, it is difficult to scale the size of the hydrogel system to proportionately increase their surface area to enhance fluid pumping

(21). Moreover, they require additional support element for making a scalable device (12, 14), this increases the cost and complexity of the device. In comparison to hydrogels, the microporous membranes suffer from an underutilization of surface area due to single channel irrigation system. Their surface area participating in evaporation at a given temperature is highly reduced, leading to low fluid pumping rate (13, 16). Therefore, the proposed micropump comprises of a radial array of branched-shaped microchannel network to irrigate the microporous paper. Such arrangement increases the fluid volume delivered to the microporous paper per unit time to maintain wetting of its entire surface, leading to an enhanced fluid pumping rate. Moreover, the branching pattern network of channels is known for providing an energy efficient fluid flow (17). In addition, cellulose based microporous paper are having better structural integrity than hydrogels after being saturated with water. Furthermore, these microchannel networks in a PDMS block integrated with a microporous paper also provided structural support to the microporous paper to prevent its collapse under its own weight during fluid flow. Such simple and scalable architecture of structurally supported free standing microporous substrate, while the micropump in operation, was unavailable in the previously described passive micropumps. The micropumps reported in literature needed an extra planner substrate onto which entire micropump assembly was mounted (13, 16). Thus, this restrains the applicability of these pumps in applications demanding any three dimensional configurations. Moreover, the PDMS material being hydrophobic in nature prevented the rapid evaporation of water from its surface analogues to the waxy cuticle present in natural leaves (Figure 1b). The PDMS material on one surface of microporous paper enabled our leaf like pump to be amiable for evaporation control via the changes in the thickness of PDMS support (22). Our calculation shows that water to permeate PDMS film of thickness 2mm takes about $6.67*10^3$ minutes, that is manifold higher than time to permeate through branched channel network(23). This arrangement facilitates that water is not lost through diffusion from PDMS surface but permeates through radially arranged branched channel network. Such control over the evaporation was only available in the previously reported micropumps through changes in the pore size of membrane which is a cumbersome process. Eventually, structural design (continuously decreasing channel diameter) of micropump permitted a unidirectional capillary flow of fluid from the reservoir to the microporous paper via the tip of the terminal veins through the assembly of branched-shaped microchannels.

The method of fabrication of micropumps presented here is a quite simple, scalable and inexpensive process, involving commonly available materials like PDMS and microporous cellulose paper. The microfabrication technique, indigenously developed in our lab, enabled the formation of branched-shaped micro-mold on a glass substrate (18). In this technique, the phenomena of Saffman-Taylor instability in Helle-Shaw cell were controlled to form a stable, repeatable branched-shaped mold of ceramic suspension

(Figure 4a) on a glass slab after heat curing at 200°c. The parameters for the fabrication of mold of ceramic suspension were optimized for micropump application. Considering the scalability of the branching pattern production by the above method, in principle, any size of the micropump can be fabricated for any applications(18). The above mold can be used for multiple times for casting radial array of branched-shaped microchannels in PDMS. PDMS offers excellent casting ability to develop negative of the mold, bond with another PDMS in a partially baked state and transparent for imaging during any flow experiments. Moreover, its availability and wide applicability in microfluidic devices deemed it to be perfect material for the choice of fabrication of micropump. The open channel surfaces are sealed by the PDMS membrane already adhered to the microporous filter paper such that only terminal branches were connected to microporous filter paper. The differences in the swelling behavior of hydrogels and bonded PDMS might produce strain at their interface, causing delamination, during micropump operation. Therefore, microporous filter paper was adopted for our micropump development. In addition, microporous filter paper, being inexpensive and readily available material, is made up of cellulose that has a high affinity for water. This supported better wicking by capillary action. The critical step in the development of the proposed micropump is an effective bonding of microporous material (here; Whattman filter paper) with non-porous material mold (PDMS) having radial arrangement of branch-shaped microchannels. Thus method of fabrication can be extended to any other type of material systems that satisfy the above conditions.

The image processing of the branched-shaped structures revealed the structural parameters of branched shaped microchannels. The depth of the microchannels showed variation across different branching generations (Figure 4a). The graph in the Figure 4b shows the variation of length and width of microchannels at different generations in a branched network. Thus, we achieved 3D multi-scale channel network commonly observed in venation system of leaves. The fractal dimensions for length and width of the branched channels network were estimated to be 0.668 and 0.754 respectively. Further, SEM of microporous filter paper shown in the Figure 4c demonstrated its morphology. The thickness of microporous paper was calculated from the inset image of figure 4c was ~170µm. The pore-size distribution of the microporous substrate were estimated from their SEM and shown in the Figure 4d. The porosity estimated by gravimetry test for the microporous support was 0.67. Further, the absorption capacity of microporous paper was evaluated by absorption studies in which it was observed that 1.082g of microporous substrate absorbed 0.965g of water in 40 second. The functional performance of leaf was gauged by fluid pumping studies.

The fluid pumping experiments with LMMs demonstrated high fluid pumping capabilities in addition to maintaining the pressure head. The experimental recordings of fluid pumping rates were carried out once the porous support of micro pumps were completely irrigated to a saturated condition. This preconditioning was carried out to negate any possibility of contribution of pumping by absorption through dry microporous substrate and bring the system to a steady state condition. At a steady state, volumetric flow through microporous substrate is equal to the evaporative flux of water from the exposed surface area of the substrate. Therefore, volumetric flow rate can directly qualify to be a measurement of fluid pumping capacity of LMMs. The measurements were performed after the steady state pumping rate was achieved by the micropump at a given pressure head and shown in the Figure 5a. Further, in a horizontal setup, when a micropump and a reservoir were placed at a same height, the average fluid pumping rate was recorded to be 0.108mg/sec. However, when the experiments were carried out with stand-alone microporous substrate connected to single channel irrigating the surface as a control, it was observed that average pumping rate was only 0.018mg/sec as shown in the Figure 5b. Although, ambient conditions, area and nature of microporous substrate participating in an evaporative pumping was same, however, an appreciable difference in pumping was recorded. This might be due to additional branched-shaped microchannel net that irrigates the microporous substrate at multiple locations at a same time as opposed to single microchannel irrigation system. This integration of branched-shaped channel network with microporous paper reduces the effective resistance to the capillary flow as compared to the porous paper alone. Moreover, multiple fluid delivery point due to branching channel might assist in keeping the microporous paper in saturated state and lead to utilization of its maximum area for evaporation.

The LMMs connected to the reservoir were slowly raised with respect to the reservoir to quantify the pressure head it could sustain before the cavitation occurs. It was observed that the fluid pumping rates of LMMs were declining at ~10 times faster than the rate of decline of the fluid pumping rate of the control microporous substrate (Figure 5a). This may be due to inability of branched-shaped microchannels to sustain pressure head above 8cm, eventually giving way for cavitation to follow by allowing air bubble to enter from the point where the terminal branches touches the microporous substrate. In the absence of smooth transition from the diameter of terminal microchannels to micropores of microporous substrate, the slight perturbation during lifting of LMM may have routed air bubble in the branched-shaped microchannels at the junction of microporous substrate and branched-shaped channels. This might have resulted in pressure loss at the interface. This loss in pressure increases with an increase in the height of pump. Nevertheless, as the diameter of the terminal branch of branched-shaped microchannel net is approximately 30 micron, the volume of fluid dispensed by branched-shaped microchannels is instantaneously absorbed by microporous substrate whose surfaces are continuously experiencing evaporation of water. This might be responsible for the potential fluid pumping mechanism exhibited by leaf-inspired micropumps. When there was continuous increase of LMM height with respect to the reservoir, it was observed that tiny air bubbles migrate continuously towards the tube connecting micropump and the reservoir. During, this process, air bubbles continuously nucleate to form large air bubble till they further prevented entry of fluid from reservoir to the LMM and thereby, release the suction pressure developed by the LMM. Thus, at the pressure head of 8cm, branched-shaped microchannel has no role in sustaining the head and the onus of sustaining the pressure head rests majorly on the microporous substrate of the micropump. This is evident from the nearly similar pumping flow rate at 8 cm pressure head for both the systems (Figure 5b). The above problem is due to limitation in the fabrication process which can be overcome by reduction in the size of terminal microchannels and making its transition nearly continuous with pore of microporous substrate. Thus, there is a distinctive role of branched-shaped microchannel net and microporous support on the performance of LMMs.

The Figure 6a summarizes the comparative fluid pumping capabilities reported by previous researchers (10-13, 16) with the LMMs proposed here. The comparison over the range of parameters shown in the Table 1 clearly illustrates the nature of the microporous substrate chosen for micropump fabrication was different from each other. The nature of microporous substrate comprises the size of micropores present in a substrate, the permeability of the substrate and degree of wettability due to hydrophobicity of the material. These factors are responsible for volumetric flow rate through these microporous substrate while the micropump in operation at a steady state condition. However, it was found that average pore-size of the substrate used for micropumps recorded in literatures were found to be around 10um with variation from 7um to 32um. In addition, the degree of hydrophobicity of materials varied from each other but primarily all the materials were hydrophilic only. Thus, the capillary pressure responsible for driving fluid through porous substrate, being the function of pore size and material property, was comparable. Further, the ambient temperature and humidity under consideration was also similar range which governs the evaporative flux of water from the surface of porous substrate. Hence, we speculate that the differences in the pumping capacity may be due differences in participation of surface area available for evaporation. Such variances over the participation of surface area for evaporation may be due to the differences in the porosity of the substrate and the diameter of the channels irrigating the surface. The hydrogels used by Wheedle et al and ling Lu et al were highly porous, hydrophilic material which accounted for their highest pumping capacity per unit area(11, 13, 14) among the two categories of micropumps. However, the proposed LMMs were showed 1.5 times higher pumping rate per unit area than the micropump developed by Robert Crawfield et al, 2013, in a category of micropumps based on microporous fibrous membranes, although the pore-size of in their membrane was about $1/4^{th}$ of pore-size of membrane available in our LMM. It is needless to say that the reduction in pore-size of the membrane not only increases the capillary pumping pressure, but also enhances the surface area of fluid for evaporation. In spite of higher pore-size in the membrane, there is enhancement in the fluid pumping rate that might be due to the presence of multiple channels irrigating the substrate and less resistance to the flow of fluid in the branched-shaped microchannels as compared to linear channels. Thus, it indicates that such strategy of branch-shaped microchannel irrigation system can be employed in conjunction with micro/nanoporous membrane to enhance the fluid pumping capacity of passive micropumps.

Finally, we have proposed the development of a simple passive micropump inspired from the leaves of plants using scalable, inexpensive fabrication method. The leaf mimicking micropump performance is dependent on its design, intrinsic and ambient parameters. Therefore, the design of the micropump should be developed considering the ambient conditions in which it would be employed. Our future work involves the investigation on the role of ambient parameters like temperature, humidity and air flow in defining the fluid pumping capacity. Nevertheless, the above micropump can still find application in number of fields from electronics to bioengineering.

## 4. Conclusion

The capillary pumping of fluid for micro/nanofluidic applications is one of most coveted areas of research. Our approaches to achieve meaningful pumping rates rest on the principle of bio-mimicking the leaves of the plants. We have designed and fabricated LMMs by a simple, scalable and inexpensive method of fabrication. Our single leaf mimicking micropump was capable of pumping fluid at the rate 388.8mg/hr and could pump the fluid to the height of 80 mm before the occurrence of cavitation. Such pumping capabilities are better than or comparable to the ones reported in the literature. We have successfully demonstrated that branched-shaped microchannel network is responsible for high pumping rate and microporous substrate for sustaining the suction head. The dependence of the pumping behaviour on the ambient temperature, humidity, number of channels irrigating the porous substrate, pore-size and permeability of the membrane need to be explored in future for further optimizing the performance of leaf inspired micropumps. Such pumping rate and suction head is demanded in micro/nanofluidic devices used for energy applications, cooling in microelectronic circuits, flow through chromatographic column etc. This study is expected to enable the development of micropumps with enhanced pumping rate without consumption of energy.

## Acknowledgement

This research effort was supported partially by Suman Mashruwala Microengineering Laboratory (www.me.iitb.ac.in/~mems) sponsored by IIT Bombay alumnus Mr. Raj Mashruwala and IITB-Monash Research Academy. The work was presented as an oral presentation in the National conference on

Convergence of Pharmaceutical Sciences and Biomedical technology, 2018, NIPER, Ahmadabad, India.

# Figures

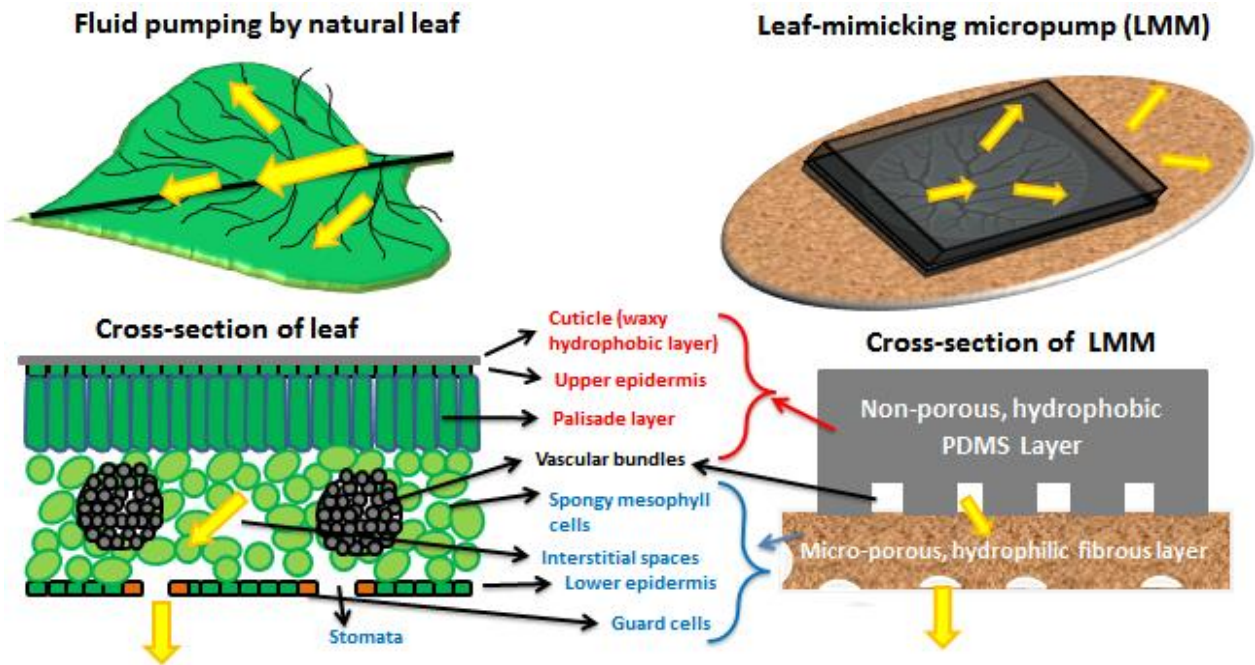

Figure 1 Schematic of water transport in trees. The yellow arrow represents the movement of water from vascular channels to an outer ambience through stomata (left leaf) and microporous support (right micropump) respectively.

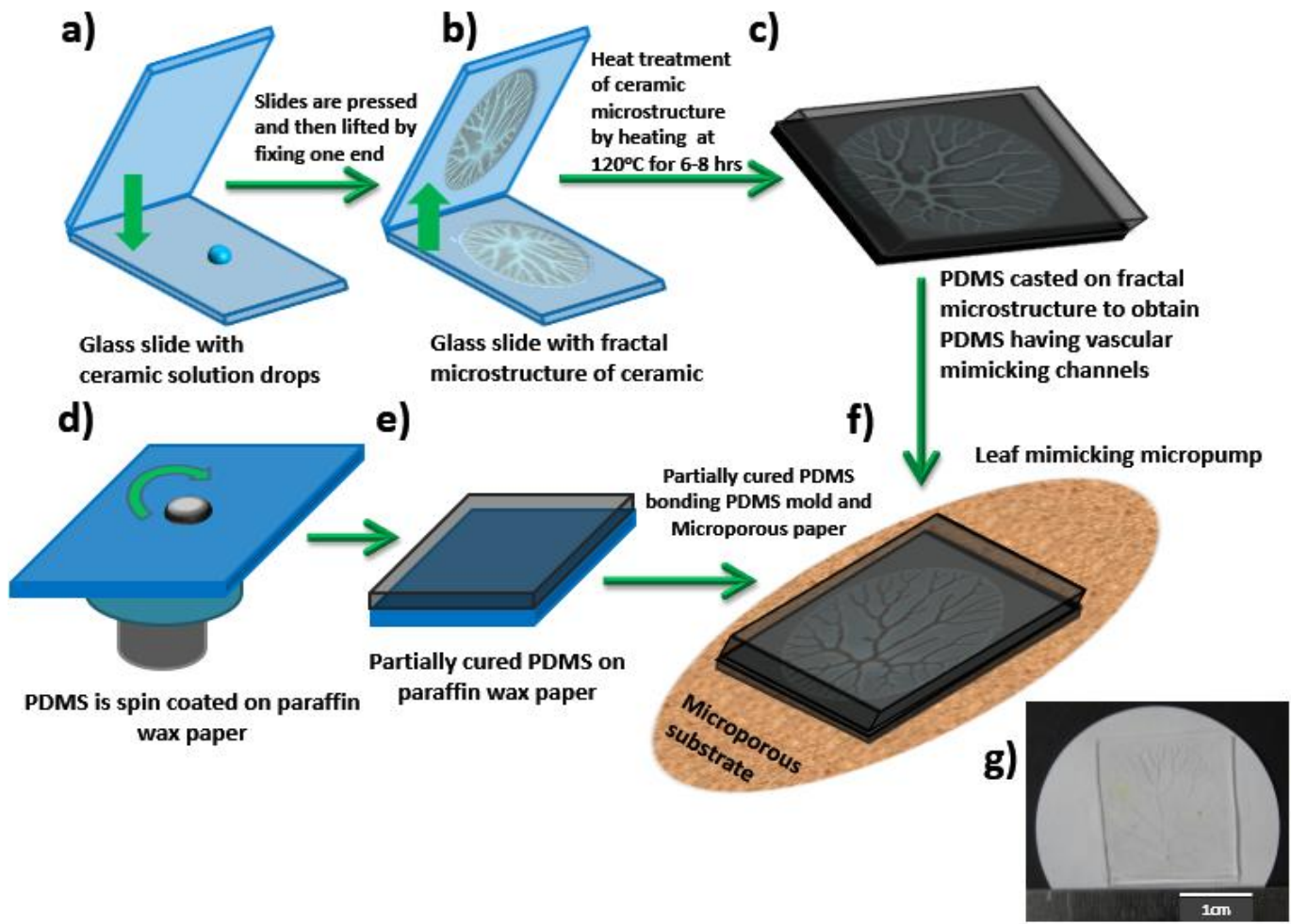

Figure 2 Schematic of fabrication process of artificial leaf in a series of steps. The steps are a) placement of a drop of ceramic suspension on a clean glass slide and sandwiching the drop to form a thin film b) angular lifting of the upper slide keeping the bottom slide fixed to form a fractal-shaped microstructure and heating it over hot plate at 120°C c) casting and curing of PDMS over the fractal-shaped ceramic mold d) spin coating of PDMS over the paraffin wax paper e) partial curing of PDMS film on paraffin paper by leaving at room temperature for 24hrs f) bonding of PDMS mold having fractal-shaped microchannel net with microporous filter paper by sandwiching a partially cured PDMS film g) Camera image of leaf-inspired micropump

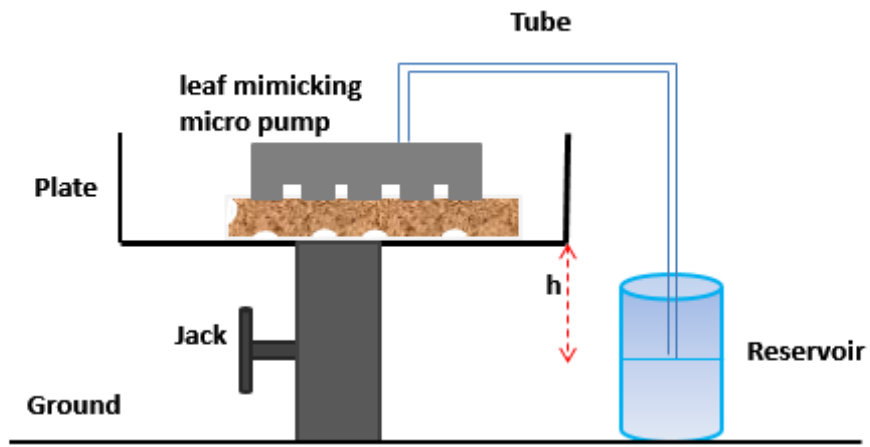

Figure 3 Schematic of the set-up to study the capillary pumping by leaf inspired micropump

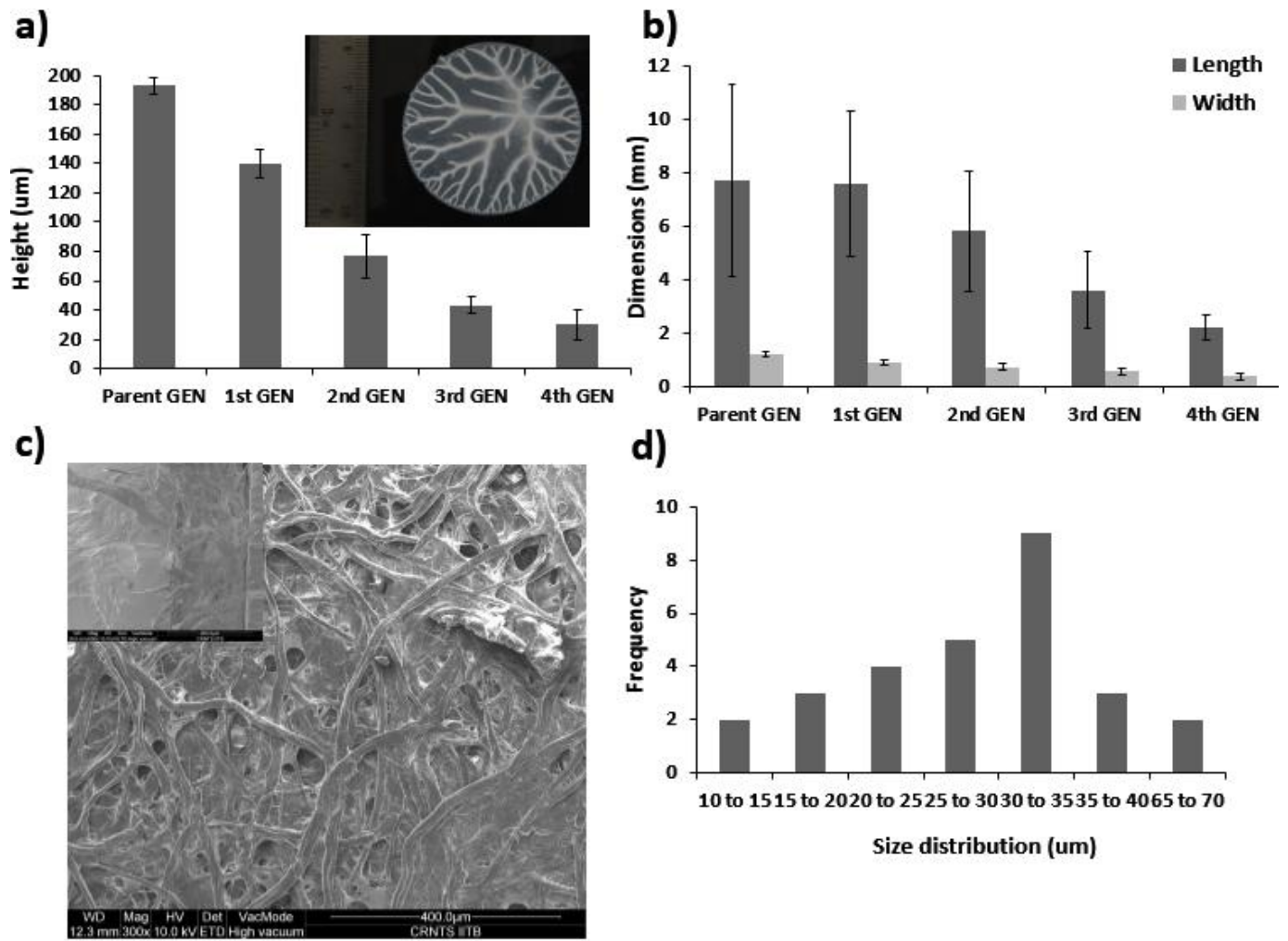

Figure 4 a) Graph showing the variation of height of fractal-shaped ceramic mold with different generations. Inset image shows fractal-shaped microstructure mold of ceramic suspension on glass slide B) Graph showing the variation of width and length of fractal-shaped mold with different generations. c) SEM of microporous structure of filter paper (inset image shows the cross-section of the sample d) Graph showing the frequency distribution of pore-size in a microporous filter paper

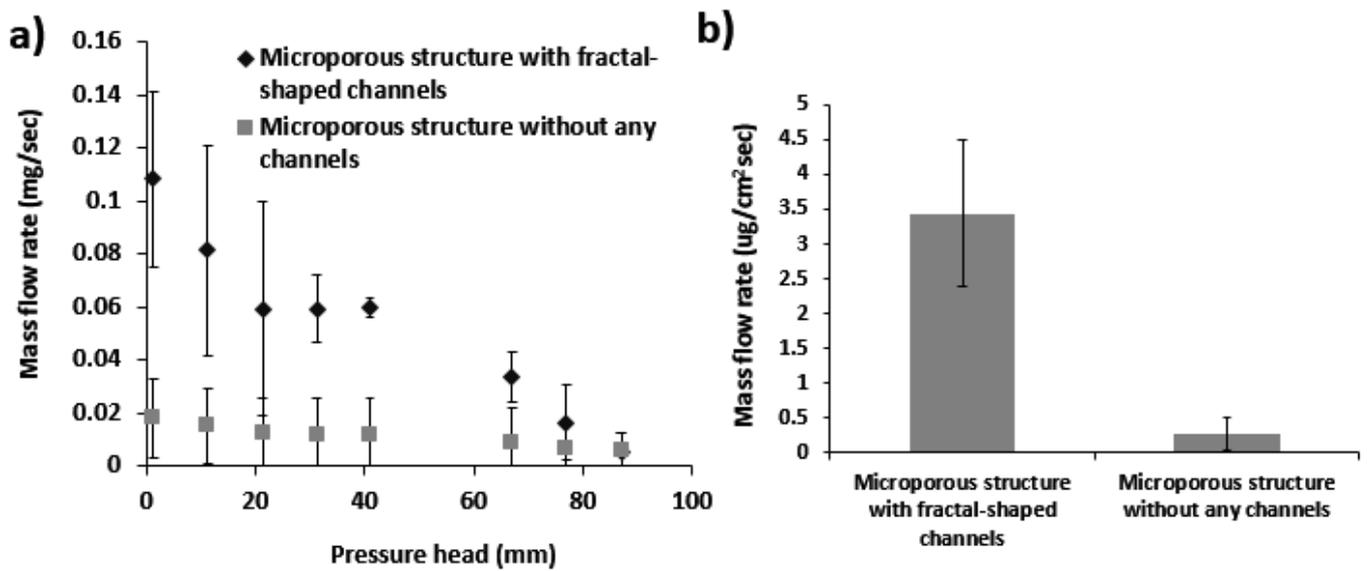

figure 5 Graph showing a) variation of mass flow rate at different suction head in microporus substrate connected with and without fractal-shaped microchannel net b) the comparison of mass flow rate achieved by microporus substrate connected with and without fractal-shaped microchannel net

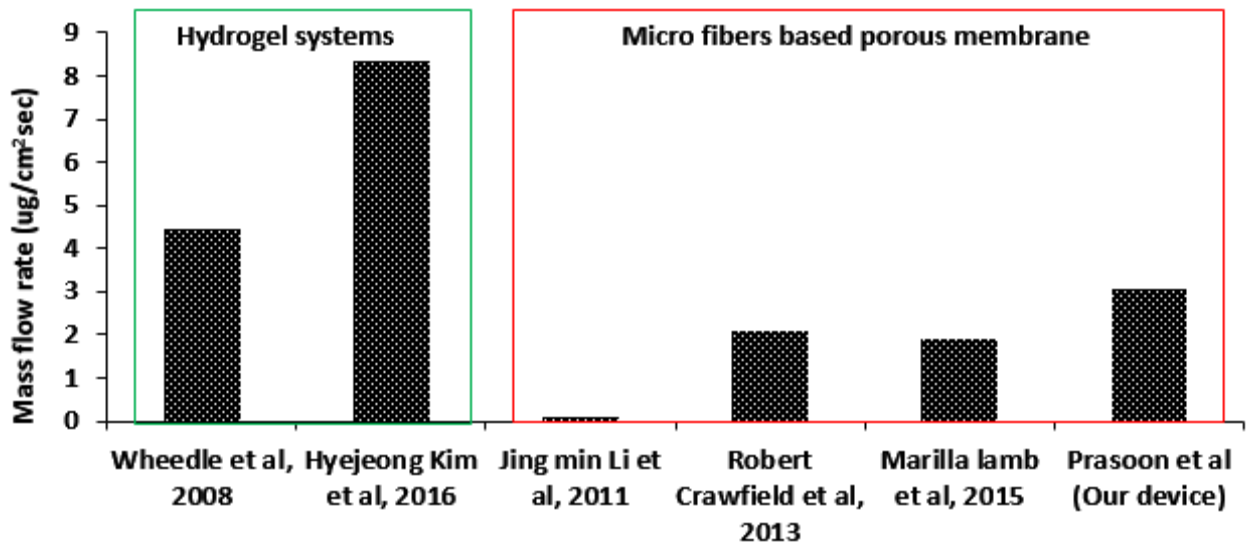

Figure 6 Graph showing the comparison of mass flow per unit area achieved by leaf mimicking micropumps developed by previous researchers with the micropump developed in the current work

Table 1 showing the differences in parameters adopted by previous researchers and current work while evaluating the pumping capacity of the micropump

|  | Type of set-up | Material | Surface area (cm2) | Temperature (K) | Pore-size (um) |
|---|---|---|---|---|---|
| Wheedle et al, 2008 | Vertical setup | PMMA hydrogel | 3.142 | 298 | 110 |
| Hyejeong Kim et al, 2016 | Vertical setup | Agarose hydrogel | _ | 298 | 106.5 |
| Jing min Li et al, 2011 | Horizontal setup | Microporous M | 0.7855 | 298 | 27 |
| Robert Crawfield et al, 2013 | Horizontal setup | Glass fibers | 176.7 | 296.5 | 7 |
| Marilla lamb et al, 2015 | Vertical setup | Ceramic | 5.4856178 | 298 | 6 |
| Prasoon et al (our device) | Vertical setup | cellulose | 35.3475 | 298 | 32 |